\documentclass[sigconf, nonacm]{acmart}

\newcommand\vldbdoi{XX.XX/XXX.XX}
\newcommand\vldbpages{XXX-XXX}
\newcommand\vldbvolume{XX}
\newcommand\vldbissue{X}
\newcommand\vldbyear{2025}
\newcommand\vldbauthors{\authors}
\newcommand\vldbtitle{\shorttitle} 
\newcommand\vldbavailabilityurl{https://github.com/yxyang1111/Pseudo-Knowledge-Graph/}
\newcommand\vldbpagestyle{plain} 

\usepackage{CJKutf8}
\usepackage{graphicx}
\usepackage{textcomp}
\usepackage{eucal}
\usepackage{xcolor}
\def\BibTeX{{\rm B\kern-.05em{\sc i\kern-.025em b}\kern-.08em
    T\kern-.1667em\lower.7ex\hbox{E}\kern-.125emX}}

\usepackage{multirow}
\usepackage{tabularx}
\usepackage{booktabs} 
\usepackage{caption}  
\usepackage{subfigure}
\usepackage{makecell}

\usepackage{balance}

\usepackage{marvosym}

\begin{document}
\begin{CJK*}{UTF8}{gbsn}

\author{Yuxin Yang}
\affiliation{%
  \institution{Peking University}
  \streetaddress{No. 5 Yiheyuan Road, Haidian District}
  \city{Beijing}
  \state{China}
}
\email{yxyang@pku.edu.cn}

\author{Haoyang Wu}
\affiliation{%
  \institution{Peking University}
  \streetaddress{No. 5 Yiheyuan Road, Haidian District}
  \city{Beijing}
  \state{China}
}
\email{why@stu.pku.edu.cn}

\author{Tao Wang}
\affiliation{%
  \institution{Peking University}
  \streetaddress{No. 5 Yiheyuan Road, Haidian District}
  \city{Beijing}
  \state{China}
}
\email{wangtao@pku.edu.cn}

\author{Jia Yang}
\affiliation{%
  \institution{Peking University}
  \streetaddress{No. 5 Yiheyuan Road, Haidian District}
  \city{Beijing}
  \state{China}
}
\email{yangj@pku.edu.cn}

\author{Hao Ma}
\affiliation{%
  \institution{Peking University}
  \streetaddress{No. 5 Yiheyuan Road, Haidian District}
  \city{Beijing}
  \state{China}
}
\email{mah@pku.edu.cn}

\author{Guojie Luo}
\affiliation{%
  \institution{Peking University}
  \streetaddress{No. 5 Yiheyuan Road, Haidian District}
  \city{Beijing}
  \state{China}
}
\email{gluo@pku.edu.cn}

\title{Pseudo-Knowledge Graph: Meta-Path Guided Retrieval and In-Graph Text for RAG-Equipped LLM
}

\begin{abstract}
The advent of Large Language Models (LLMs) has revolutionized natural language processing. However, these models face challenges in retrieving precise information from vast datasets. Retrieval-Augmented Generation (RAG) was developed to combining LLMs with external information retrieval systems to enhance the accuracy and context of responses. Despite improvements, RAG still struggles with comprehensive retrieval in high-volume, low-information-density databases and lacks relational awareness, leading to fragmented answers.

To address this, this paper introduces the Pseudo-Knowledge Graph (PKG) framework, designed to overcome these limitations by integrating Meta-path Retrieval, In-graph Text and Vector Retrieval into LLMs. By preserving natural language text and leveraging various retrieval techniques, the PKG offers a richer knowledge representation and improves accuracy in information retrieval. Extensive evaluations using Open Compass and MultiHop-RAG benchmarks demonstrate the framework's effectiveness in managing large volumes of data and complex relationships.

\end{abstract}


\maketitle

\pagestyle{\vldbpagestyle}
\begingroup\small\noindent\raggedright\textbf{PVLDB Reference Format:}\\
\vldbauthors. \vldbtitle. PVLDB, \vldbvolume(\vldbissue): \vldbpages, \vldbyear.\\
\href{https://doi.org/\vldbdoi}{doi:\vldbdoi}
\endgroup
\begingroup
\renewcommand\thefootnote{}\footnote{\noindent
This work is licensed under the Creative Commons BY-NC-ND 4.0 International License. Visit \url{https://creativecommons.org/licenses/by-nc-nd/4.0/} to view a copy of this license. For any use beyond those covered by this license, obtain permission by emailing \href{mailto:info@vldb.org}{info@vldb.org}. Copyright is held by the owner/author(s). Publication rights licensed to the VLDB Endowment. \\
\raggedright Proceedings of the VLDB Endowment, Vol. \vldbvolume, No. \vldbissue\ %
ISSN 2150-8097. \\
\href{https://doi.org/\vldbdoi}{doi:\vldbdoi} \\
}\addtocounter{footnote}{-1}\endgroup

\ifdefempty{\vldbavailabilityurl}{}{
\vspace{.3cm}
\begingroup\small\noindent\raggedright\textbf{PVLDB Artifact Availability:}\\
The source code, data, and/or other artifacts have been made available at \url{\vldbavailabilityurl}.
\endgroup
}

\balance

\section{Introduction}\label{sec:intro}

The emergence of large language models (LLMs)~\cite{radford2019language, brown2020language} has transformed natural language processing, allowing machines to understand and generate text that closely resembles human communication~\cite{wei2022emergent}. These models, trained on extensive datasets, excel in various applications \cite{wang2023pre}, including chatbots and content creation. However, despite their capabilities, LLMs encounter significant challenges~\cite{zhai2008statistical} when tasked with retrieving specific information from extensive collections of data. This often results in incomplete or imprecise answers, particularly when users seek detailed insights~\cite{hadi2023survey, tamkin2021understanding}. Despite the increasing capabilities of LLMs, deploying them with private data and ensuring the authenticity of generated text remain significant challenges. Fine-tuning LLMs for specific domains and managing private data incur high costs, especially when base models are frequently updated, requiring repeated fine-tuning. Additionally, LLMs cannot inherently verify the truthfulness of their outputs, necessitating the extraction of third-party facts to support their claims. To mitigate these limitations~\cite{burtsev2023working}, researchers have developed Retrieval-Augmented Generation (RAG)~\cite{lewis2020retrieval}, a hybrid approach that combines LLMs with external information retrieval systems. RAG addresses these issues by enabling LLMs to retrieve and reference external data, enhancing both the accuracy and authenticity of generated responses~\cite{siriwardhana2023improving}.

Nevertheless, RAG is not a universal solution~\cite{bruckhaus2024rag}. \emph{One of its fundamental shortcomings arises when the information needed is scattered across a vast knowledge base, creating challenges in retrieving comprehensive answers.} This issue is particularly pronounced in large databases characterized by low information density, high redundancy, and dispersed information~\cite{cuconasu2024power}. \emph{Additionally, traditional RAG systems often struggle to discern and leverage the relationships between different pieces of information.} From the perspective of authenticity, relying solely on the top-1 or top-3 results based on similarity metrics in vector databases is often insufficient~\cite{gao2023precise}. Multiple supporting facts are required to ensure the reliability of retrieved information. While vector databases excel at retrieving semantically similar items, they lack mechanisms to ensure diversified proximity, which is crucial for capturing complex relationships. Complex relationships, such as multi-hop connections or indirect associations between entities, cannot be adequately represented by simple similarity metrics\cite{yang2018hotpotqa}. This limitation underscores the need for more sophisticated retrieval methods, such as meta-path-based approaches, which can uncover intricate relational pathways and provide a richer context for LLMs~\cite{pan2024vector}.

To address these challenges, there is a pressing need for innovative storage and retrieval methods that can harness the strengths of vector databases while overcoming their limitations. Traditional approaches that integrate LLMs with knowledge graphs (LLM-KG) leverage the structured nature of graphs to provide contextual relationships and factual grounding, improving the accuracy and coherence of generated responses. However, these systems face significant limitations. LLMs often struggle to process structured graph data effectively, leading to incomplete or fragmented answers~\cite{sui2024table, meyer2023llm}. Additionally, traditional knowledge graphs are static and may not capture the dynamic nature of real-world knowledge, while their integration with LLMs typically requires extensive fine-tuning and domain-specific adaptation, which is computationally expensive~\cite{pan2023large}. These challenges highlight the need for a more flexible and scalable approach that bridges the gap between structured and unstructured data.

This paper introduces the Pseudo-Knowledge Graph (PKG), an innovative framework that enhances the processing of \emph{large-scale} information by addressing challenges related to \emph{complex} data relationships. Building on the RAG paradigm, PKG integrates knowledge graphs, meta-path retrieval, and natural language text preservation to create a robust and context-aware retrieval system. At its core, PKG stores structured representations of entities and their relationships while preserving the original text chunks, enabling LLMs to process and interpret information effectively, overcoming their limitations with purely structured data. PKG employs advanced retrieval techniques, including vector-based retrieval for semantic similarity and meta-path retrieval for uncovering complex, multi-hop relationships (e.g., "author-paper-conference" or "disease-symptom-treatment"). These methods allow PKG to identify semantically relevant information and explore intricate relational pathways, fostering a deeper understanding of context and connections. By seamlessly integrating structured and unstructured data, PKG excels in scenarios requiring multi-hop reasoning and contextual awareness, such as scientific research, legal analysis, and healthcare. This approach improves the accuracy and relevance of generated answers, empowering users to navigate complex knowledge bases effectively and make more informed decisions.


To assess the effectiveness of our method, we employed several widely used large language models to generate a diverse array of questions based on two benchmarks: Open Compass and MultiHop-RAG. This approach allowed us to thoroughly examine the performance of our framework across different scenarios and contexts. The contributions of this work can be summarized as follows:

\begin{itemize}
    \item We present a framework for constructing and retrieving knowledge in the form of Pseudo-Knowledge Graph (PKG). This framework enables language models to accurately retrieve relevant information from a vast array of discrete knowledge.
    
    \item We integrate multiple retrieval techniques into the PKG search, including regular expression matching, vector retrieval, relation-based retrieval, and meta-path retrieval, yielding strong results in information retrieval.
    
    \item We conducted extensive evaluations based on the Open Compass and MultiHop-RAG benchmarks across multiple commonly used models, demonstrating the superior performance of the PKG framework in handling large volumes of information and complex relationships in the knowledge base.

\end{itemize}

\section{Related Work}

\subsection{Retrieval Augmented Generation}

Shortly after the introduction of pre-trained language models \cite{kenton2019bert}, Large Language Models (LLMs) \cite{brown2020language} have significantly advanced natural language processing, excelling in tasks like translation and summarization \cite{chang2024survey, zhao2023survey}. However, they often struggle with factual accuracy, generating outdated or incorrect information due to reliance on learned patterns. To address these challenges, the Retrieval-Augmented Generation (RAG) framework~\cite{lewis2020retrieval} is introduced. RAG enhances the generative capabilities of LLMs by incorporating a retrieval mechanism that accesses relevant information from external knowledge bases~\cite{li2022survey}. This two-step process retrieves relevant documents based on the input query and uses them to inform response generation. By integrating retrieval and generation, RAG enhances factual accuracy and enriches content with current, contextually relevant information \cite{jiang2023active}.

RAG has shown promising results in applications like question-answering and conversational agents, setting a new standard for combining retrieval and generative techniques. The integration of vector databases with RAG holds significant potential for improving the efficiency and effectiveness of information retrieval alongside LLMs \cite{salemi2024evaluating}.

\subsection{Knowledge Graph}

Before the advent of LLMs, Knowledge Graphs (KGs) \cite{fensel2020introduction} were a preferred choice for information retrieval \cite{reinanda2020knowledge} and smart Q\&A \cite{yasunaga2021qa, zou2020survey}. KGs are structured representations of knowledge, capturing relationships between entities in a graph format. They consist of nodes (entities) and edges (relations), organizing information in a machine-readable and human-understandable way. This framework integrates diverse data sources, representing complex relationships and concepts within a unified structure, often enriched with metadata for enhanced contextual understanding.

KGs have diverse applications \cite{zou2020survey, hao2021construction}. In search engines, they improve relevance by providing contextual information about entities. In natural language processing, they enhance question-answering systems by linking queries to relevant knowledge. KGs also aid recommendation systems by understanding user preferences. Industries like healthcare, finance, and e-commerce use KGs for data integration and decision support, driving innovation and efficiency across various domains.

\subsection{Interaction between Language Models and Knowledge Graphs}

Integrating language models with knowledge graphs is crucial for advancing natural language processing \cite{pan2024unifying,pan2023large}. Language models are adept at understanding and generating human-like text, offering flexibility and contextual awareness \cite{kaddour2023challenges,hadi2023survey}. In contrast, knowledge graphs provide structured information, capturing relationships and facts to ensure accuracy and coherence.

This synergy allows language models to help build knowledge graphs by identifying entities and relationships in unstructured text \cite{meyer2023llm,zhang2024extract}. Conversely, knowledge graphs enhance language models by incorporating structured knowledge into training and inference \cite{abu2024knowledge,shu2024knowledge}, improving text accuracy and reasoning capabilities.

During pre-training, knowledge graph triples can be converted into text to help language models learn structured information, improving their understanding of factual knowledge \cite{zhang2024knowledge}. For example, models like ERNIE 3.0 \cite{sun2021ernie} use tokenized triples to mask entities and relationships, promoting effective learning. During inference, language models retrieve real-time information from knowledge graphs, enabling precise and contextually relevant responses.

Moreover, language models are essential for updating knowledge graphs by extracting new entities and relationships from recent data \cite{edge2024local}. This ongoing process ensures the graphs remain accurate and relevant. By detecting inconsistencies and suggesting updates, language models significantly enhance the quality of knowledge representation, improving the effectiveness of natural language processing applications.

\section{Methodology}\label{sec:methodology}

\begin{figure}[t]
    \centering
    \includegraphics[width=0.45\textwidth]{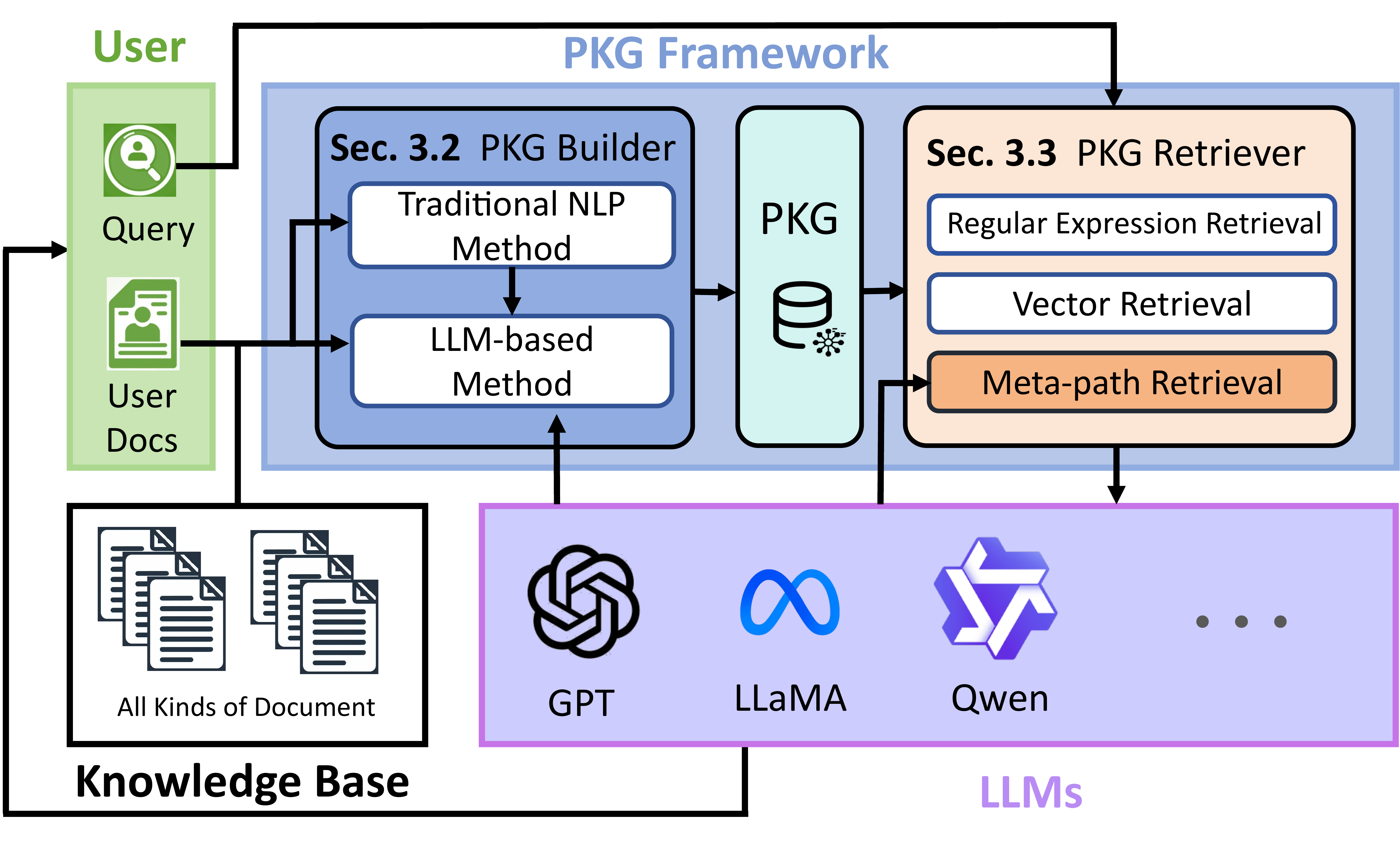}
    \caption{The overall framework of our PKG approach. We enhance LLMs by integrating diverse methods for building and retrieving PKG. }
    \label{fig:framework}
\end{figure}

In this section, we present the proposed Pseudo-Knowledge Graph (PKG)-based information retrieval system, which integrates PKG semantics and collaborative semantics to enhance the performance of large language models (LLMs). The overall framework of the proposed PKG approach is illustrated in Figure~\ref{fig:framework}.

The PKG is a sophisticated RAG framework designed to tackle the challenges of processing vast amounts of information and managing intricate data relationships. PKG leverages the complementary strengths of knowledge graphs, LLMs, and meta-path retrieval to build a highly adaptable and context-sensitive retrieval system. Central to its design, PKG stores diverse representations of entities and their interconnections within a structured graph framework, while also retaining the original natural language text segments from which these elements are derived. This hybrid approach—combining structured graph data with unstructured text—ensures that LLMs can efficiently interpret and utilize the retrieved information, bypassing their typical difficulties with purely structured data formats. By preserving the richness of natural language, PKG enhances the LLMs' ability to generate accurate and contextually relevant responses, making it a powerful tool for navigating complex knowledge domains.

\subsection{Overview of the Approach}

As discussed in Section~\ref{sec:intro}, traditional RAG systems relying on vector databases struggle to effectively manage large volumes of complex information~\cite{zhao2024chat2data}, which significantly limits the capabilities of LLMs when processing extensive knowledge bases. To address these limitations, we propose the Pseudo-Knowledge Graph (PKG) framework, a novel approach designed to enhance semantic understanding, relation extraction, and information retrieval efficiency. PKG achieves this by integrating structured knowledge representations with unstructured natural language text, enabling LLMs to process and interpret complex data more effectively. The framework consists of two core components:

\begin{itemize}
    \item The PKG Builder (Section~\ref{sec:PKG builder}) is an automated tool for constructing PKGs. It employs advanced techniques to identify entities and extract relationships from unstructured text, transforming raw data into a structured graph format. By combining traditional NLP algorithms (e.g., tokenization, dependency parsing) with state-of-the-art language model techniques, the PKG Builder ensures high accuracy and scalability in graph construction. This hybrid approach leverages the strengths of both rule-based methods and machine learning models, resulting in a more reliable and precise representation of knowledge compared to existing methods. The PKG Builder also preserves the original text chunks within the graph, enabling LLMs to process information in its natural language form, thereby overcoming their limitations in handling purely structured data.
    
    \item The PKG Retriever (Section~\ref{sec:PKG Retriever}) facilitates efficient and flexible information retrieval from the constructed PKG. It supports a variety of retrieval methods, including keyword searches, semantic searches, and meta-path searches, enabling users to execute complex queries that leverage the relationships and attributes defined in the PKG. The PKG Retriever is designed with a user-centric interface, allowing users to filter results, visualize entity connections, and extract actionable insights. By combining these capabilities, the PKG Retriever enhances the decision-making process for LLMs, enabling them to generate more accurate and contextually relevant responses. The integration of meta-path retrieval, in particular, allows for the exploration of complex, multi-hop relationships, which is critical for tasks requiring deep contextual understanding.

\end{itemize}

In the following sections, we provide a detailed explanation of the methodology, including the construction of the PKG and the retrieval mechanisms that underpin its effectiveness.

\begin{figure*}[t]
    \centering
    \includegraphics[width=0.95\textwidth]{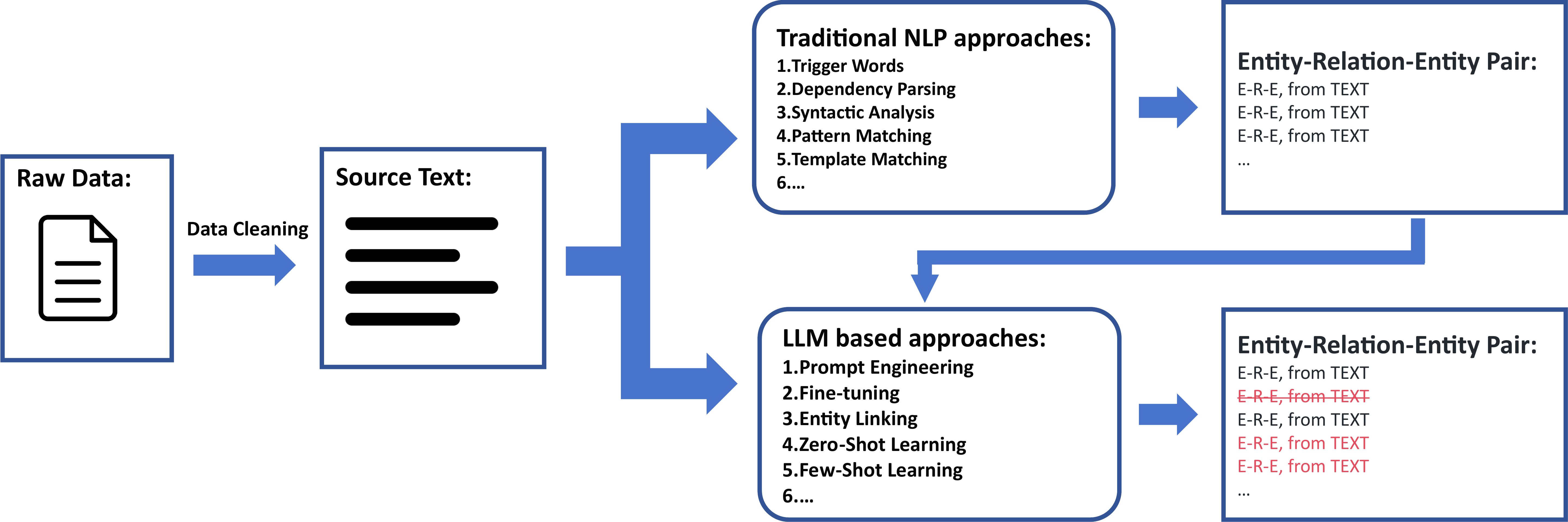}
    \caption{The extraction of entities and relations in PKG Builder. After transformation raw data into source text, We use two distinct approaches: traditional methods utilizing NLP approaches and modern techniques employing LLMs. Also, we employ LLMs to review and verify the information extracted using traditional NLP methods.}
    \label{fig:builder}
\end{figure*}

\subsection{PKG Builder}\label{sec:PKG builder}

To construct the PKG, a key challenge is accurately extracting and representing entities and relationships from unstructured text. We adopt a hybrid approach that integrates traditional NLP algorithms with advanced language model techniques, enhancing entity recognition and relation extraction. This section outlines the PKG Builder's methodology, which consists of two main steps: (1) applying NLP algorithms to identify entities and relations, converting raw data into a structured format; and (2) refining the extraction process using language models. Additionally, we optimize storage methods to improve data accessibility, scalability, and flexibility. The overall process is illustrated in Figure~\ref{fig:builder}.

\subsubsection{Entity and Relation Extraction with NLP methods}

Entity and relation extraction is fundamental to PKG construction, involving the identification of entities (e.g., people, organizations, locations) and their relationships. To achieve high performance, we integrate multiple methods to automate this process.

The extraction pipeline begins with text segmentation and compression. Text is broken into manageable units, such as sentences and phrases, using methods like sentence boundary detection and tokenization~\cite{palmer2000tokenisation, minixhofer2023s}. For compression, techniques like summarization and noise reduction are applied to remove redundant or irrelevant information, improving efficiency. Extractive summarization~\cite{zhong2020extractive, liu2019fine} and stop-word removal~\cite{silva2003importance, raulji2016stop} streamline the text, laying the groundwork for accurate entity and relation extraction.

For entity extraction, we use traditional NLP methods, including handcrafted rules, regular expressions, and linguistic cues, which are precise in well-defined contexts but require domain-specific knowledge. We also employ models like Conditional Random Fields (CRFs)~\cite{peng2006information, patil2020named} and Hidden Markov Models (HMMs)~\cite{sarawagi2004semi, scheffer2001active}, which incorporate features such as part-of-speech tags, capitalization, and contextual information.

For relation extraction, we use syntactic parsing, particularly dependency parsing, to analyze sentence structures and identify potential relationships. Rule-based patterns, defined using syntactic structures or specific phrases, are applied to detect relations. Additionally, machine learning models such as Support Vector Machines (SVMs)~\cite{hong2005relation} and decision trees~\cite{sato2001rule, yang2006extracting} are employed to classify relationships, leveraging labeled datasets and features like word pairs and dependency paths~\cite{washio2018filling}. Statistical co-occurrence measures are also used to infer relationships based on the frequency of entity co-occurrences. This hybrid approach ensures robust and accurate entity and relation extraction.

\subsubsection{Entity and Relation Extraction with LLMs}

To further enhance extraction, we incorporate LLMs, as outlined in Section~\ref{sec:models}. Using a multipart prompt, we first identify entities, detailing their names, types, and descriptions, and then discern relationships between them, specifying source and target entities. The extracted data is consolidated into a list of delimited tuples. To tailor LLM performance to specific domains, we employ few-shot learning~\cite{wang2020generalizing, song2023comprehensive}, which is particularly effective in specialized fields like science, medicine, and law. The default prompt captures a broad range of entities but can be customized with domain-specific examples for improved precision.

To ensure completeness and quality, we implement a multi-round gleaning process. The LLM first assesses whether all entities have been extracted, using a logit bias for binary decisions. If missing entities are detected, a continuation prompt triggers the LLM to recover them, ensuring high-quality extraction even for large text chunks. This iterative approach minimizes noise while maximizing data completeness.

Additionally, LLMs are used to review and verify information extracted by traditional NLP methods. By combining insights from both approaches, we achieve a comprehensive and accurate final result. This includes capturing relevant claims associated with entities, such as subject, object, type, description, and temporal information, enhancing the depth and precision of the extracted data.

By integrating LLMs with traditional NLP techniques, the PKG Builder achieves a robust and scalable solution for entity and relation extraction, tailored to the specific requirements of diverse domains.

\begin{figure}
    \centering
    \subfigure[Node Properties]{
        \raisebox{1.5cm}{\includegraphics[width=0.15\textwidth]{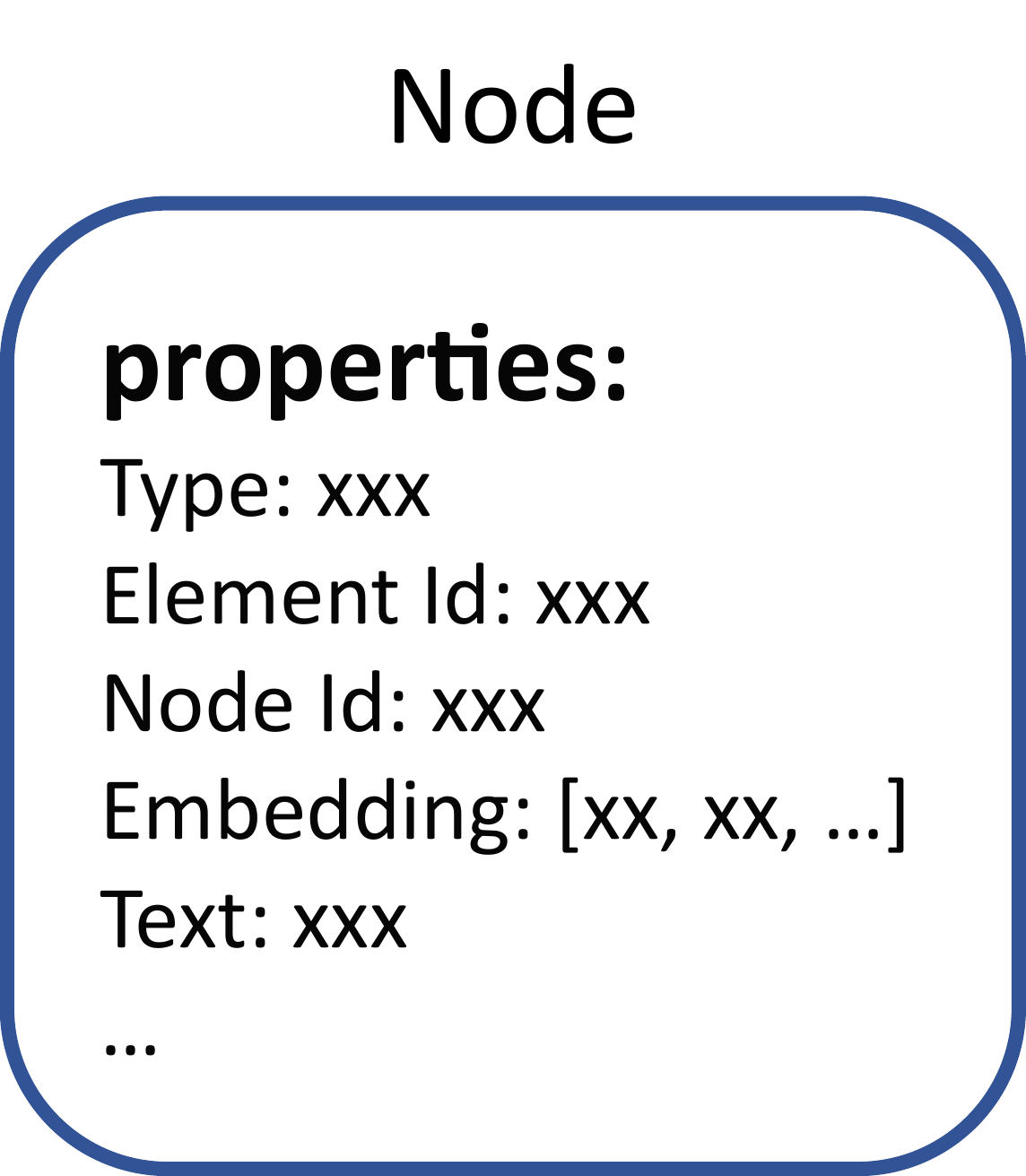}}
        \label{fig:vector sub1}
    }
    \subfigure[Relations Between Nodes]{
        \includegraphics[width=0.25\textwidth]{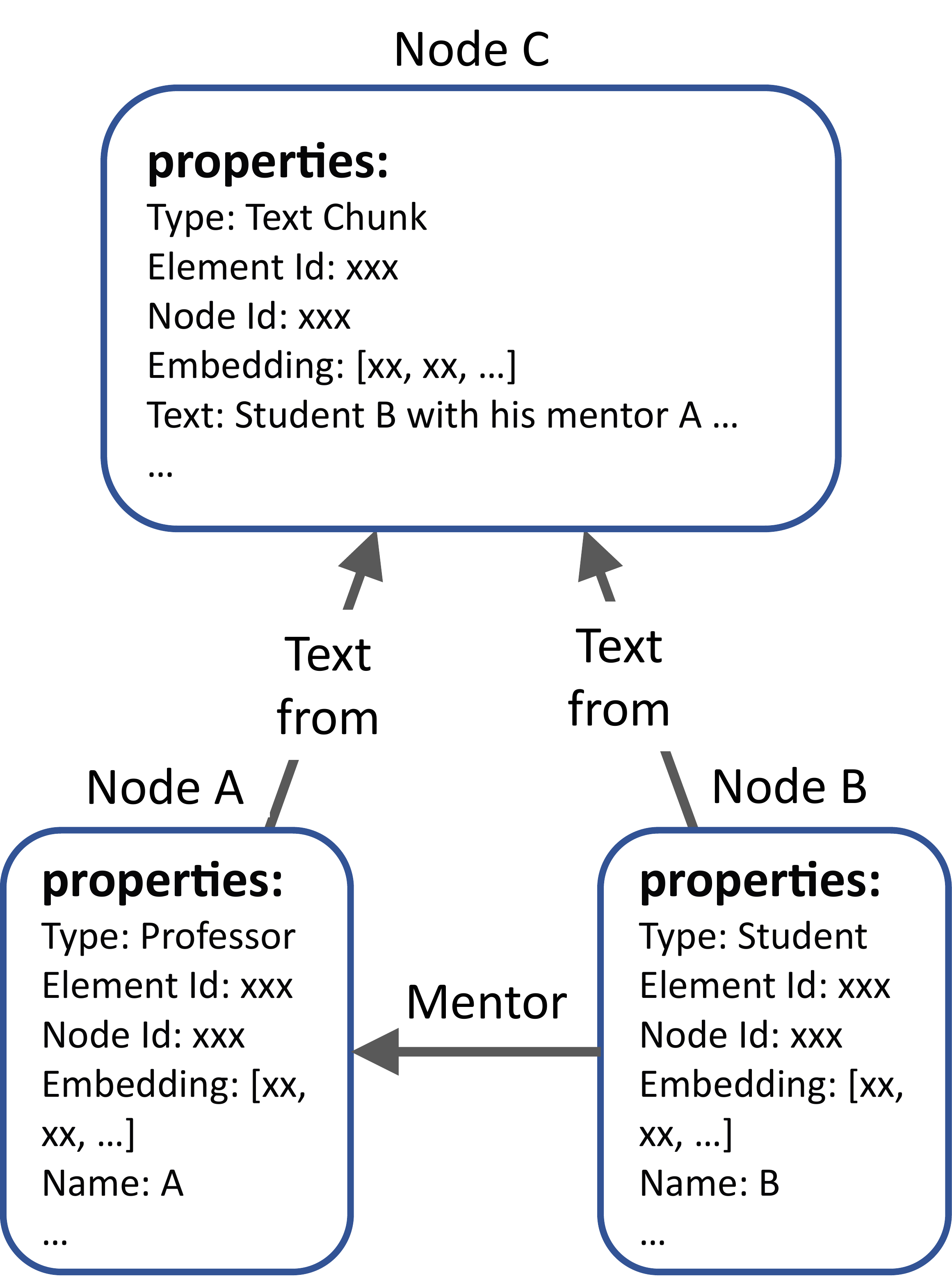}
        \label{fig:vector sub2}
    }
    \caption{Nodes and Their Properties. (a) illustrates the components of a basic node; (b) presents an example of two entity nodes extracted from a single text chunk node.}
    \label{fig:vector storage}
\end{figure}

\begin{figure}
    \centering
    \includegraphics[width=0.45\textwidth]{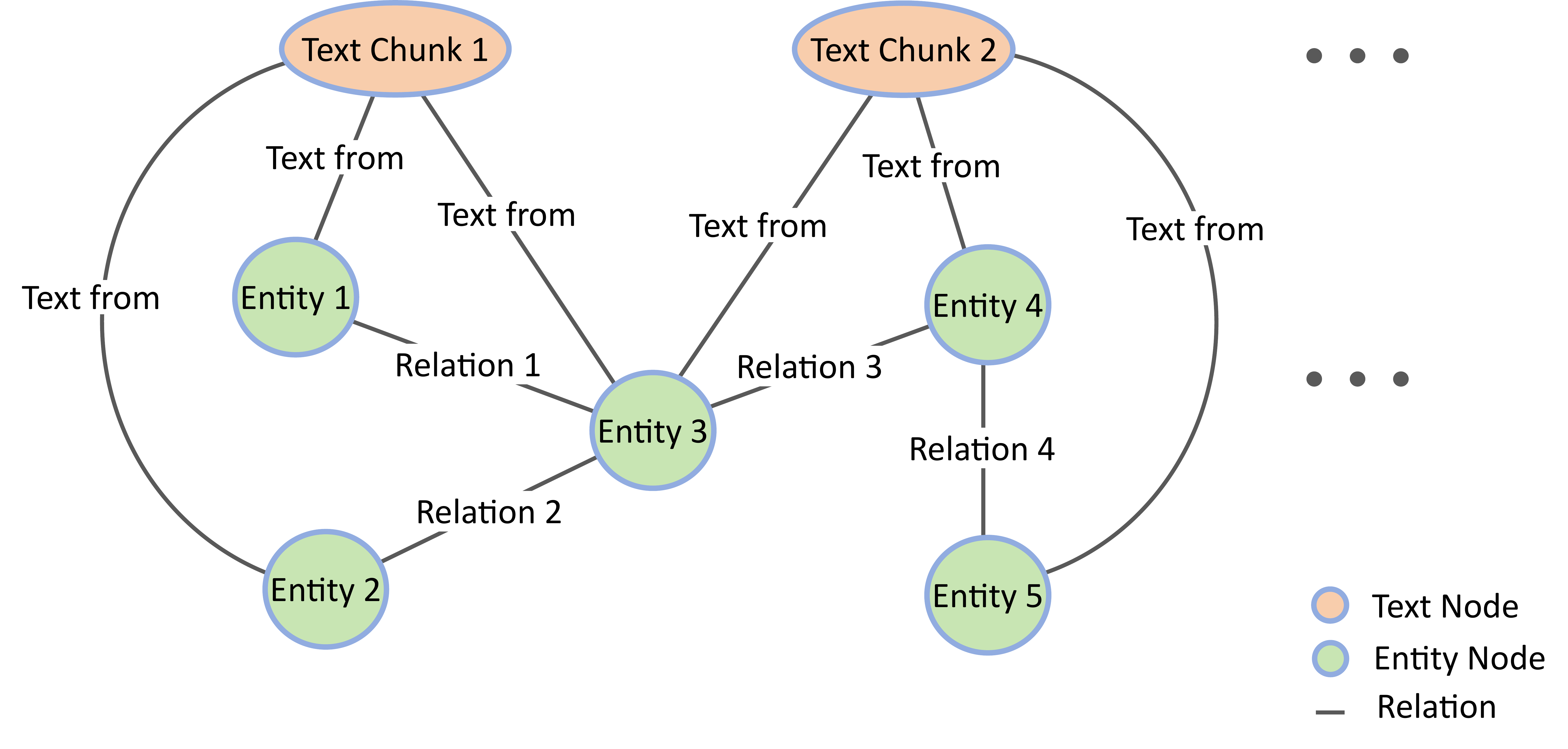}
    \caption{The organization of text data within a PKG Storage System. Each entity node must be connected to at least one source text chunk node.}
    \label{fig:storage}
\end{figure}

\subsubsection{Storage of Entities and Relations}

After extracting entities and relationships from text-based knowledge bases, effective storage is crucial for efficient querying. While traditional knowledge graphs~\cite{shu2024knowledge, edge2024local} excel in managing structured data and complex relationships~\cite{kejriwal2022knowledge}, they often fall short in supporting LLMs, which perform better with natural language~\cite{sui2024table, zhang2024applications}. To address this, we develop an optimized storage structure that combines the strengths of graph databases and natural language text.

We use graph databases like Neo4j~\cite{miller2013graph, guia2017graph} and OrientDB~\cite{ritter2021orientdb, tesoriero2013getting} to store the PKG, as they efficiently handle complex relationships. Entities and their attributes are stored as nodes, and relationships are represented as edges. To enhance query speed and semantic analysis, we vectorize each node using techniques like Word2Vec~\cite{church2017word2vec}, GloVe~\cite{pennington2014glove}, or transformer-based models like BERT~\cite{kenton2019bert}. These vectors capture the semantic meaning of nodes in a high-dimensional space, enabling fast similarity searches (e.g., cosine similarity) for efficient retrieval of related concepts or entities, as shown in Figure~\ref{fig:vector storage}.

A key innovation in our approach is the integration of in-graph text. Unlike traditional knowledge graphs that rely solely on structured data, we store the original segmented text chunks as nodes in the graph database, linking them to the corresponding entities (see Figure~\ref{fig:storage}). This ensures that during queries, relevant natural language text passages can be provided to LLMs, leveraging their strength in processing unstructured text. For example, in legal document analysis, linking case law text segments to specific legal entities allows LLMs to interpret nuanced legal language more effectively. Similarly, in scientific research, associating text from research papers with scientific concepts enables precise retrieval and understanding of complex topics.

By combining graph databases, vectorization, and in-graph text, we create a robust storage system that leverages both structured and unstructured data. Graph databases handle complex relationships, vectorization enables fast semantic searches, and in-graph text enhances LLMs' ability to process natural language, ensuring accurate and context-rich responses. This hybrid approach addresses the limitations of traditional RAG and knowledge graph systems, making PKG a powerful tool for knowledge retrieval and reasoning.

\begin{figure*}[ht]
    \centering
    \includegraphics[width=0.95\textwidth]{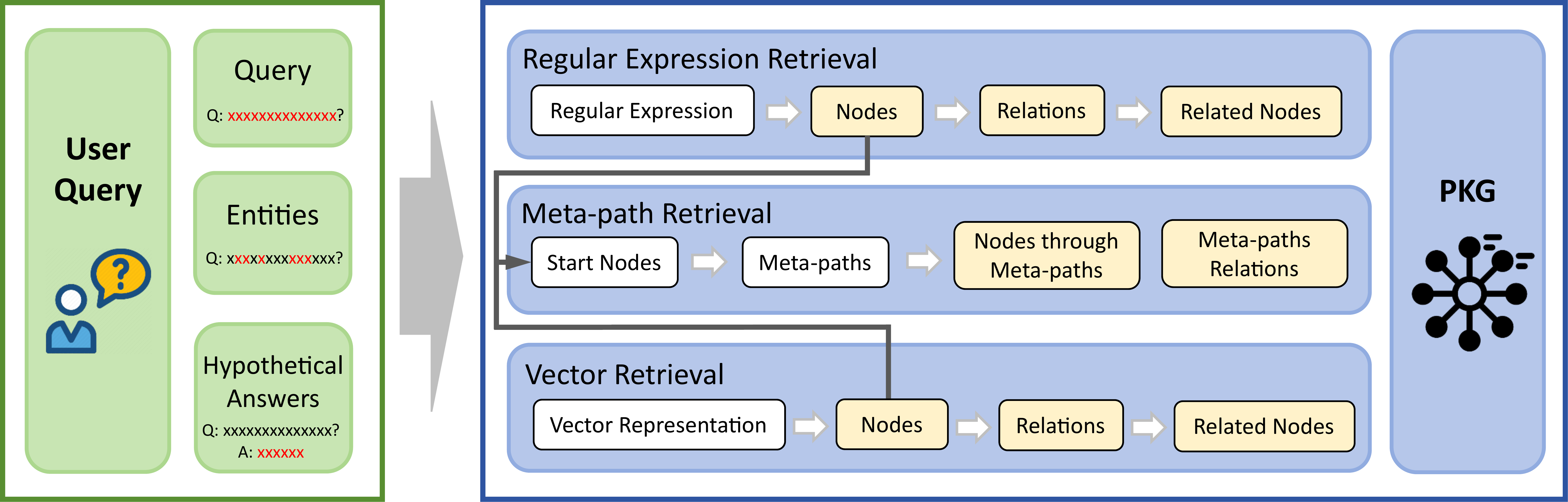}
    \caption{PKG Retriever. The retrieval process begins with a user query. Then, we get the query itself, entities inside the query, and hypothetical answers for retrieval. The retrieval methods are categorized into three types: Regular Expression Retrieval, which utilizes regular expressions to identify nodes and their relations; Vector Retrieval, which employs vector-based methods to find relevant nodes and their associated relations; and Meta-path Retrieval, which explores start nodes and their connections through specified meta-paths. The content in the light yellow boxes is what we can obtain from the PKG Retriever.}
    \label{fig:retriever}
\end{figure*}

In summary, The PKG Builder enhances the construction of PKG by integrating advanced language models with traditional NLP techniques. Entities and relations are identified using a combination of NLP methods and then refined through language models to ensure accuracy and completeness. The extracted data is stored in graph databases, optimized for efficient retrieval, and enhanced by vectorization techniques. This allows for seamless integration of structured and unstructured data, facilitating more effective querying and utilization by language models. The resulting system is scalable and flexible, supporting complex data interactions and retrieval across diverse domains, such as legal analysis, scientific research, and healthcare.

\subsection{PKG Retriever}\label{sec:PKG Retriever}

Given a user query, we can extract a wealth of information, including the query itself, its vector representation (capturing semantic information), the entities it contains, the relations required to answer it, and hypothetical answers.  Leveraging this diverse information, we develop three retrieval methods to access the prepared PKG: regular expression matching, vector-based retrieval, and meta-path retrieval. Each method exploits distinct aspects of the query-derived information, enabling efficient and effective identification of the most relevant data. Regular expression matching is used for precise pattern-based searches, vector-based retrieval leverages semantic similarity for flexible matching, and meta-path retrieval uncovers complex relational pathways between entities. By combining these techniques, we ensure a comprehensive and adaptable retrieval process capable of handling queries of varying complexity and specificity. This multi-layered strategy enhances retrieval performance and ensures robustness across diverse query types. The overall framework of the PKG Retriever is illustrated in Figure~\ref{fig:retriever}.

\subsubsection{Regular Expression Retrieval}

Regular expression retrieval is a straightforward yet effective method for extracting information from a predefined dataset or text corpus. It excels at handling queries involving specific entities or patterns by matching predefined string patterns within the PKG. For example, a regular expression can be designed to identify common date formats (e.g., ``YYYY-MM-DD'' or ``DD/MM/YYYY.'') in a document, enabling precise extraction of relevant information. This method is particularly useful in PKGs, where entities and their relationships are stored in a structured format, allowing for efficient pattern-based searches.

When a node is retrieved using a regular expression, it provides access to a cluster of interconnected nodes and their associated information. This capability is essential for tasks requiring contextual understanding, such as extracting event sequences from timelines or identifying relationships between entities. For instance, in a bibliographic PKG, regular expressions can retrieve nodes containing publication years within a specified range, facilitating the extraction of relevant articles or papers.

While regular expression retrieval is powerful for structured data, it can be combined with other retrieval techniques (e.g., vector-based or meta-path retrieval) to enhance its effectiveness, as discussed in Section~\ref{sec:Post Processing}. In summary, regular expression retrieval provides a robust mechanism for accessing structured information through pattern recognition, serving as a foundational method for information retrieval in PKGs.

\subsubsection{Vector Retrieval}

Vector retrieval is an advanced method for extracting information from the PKG by leveraging vector space models. Unlike regular expression retrieval, which relies on exact pattern matching, vector retrieval identifies semantically similar nodes by embedding entities and their contextual relationships into a high-dimensional vector space. This enables the retrieval of nodes that share semantic relevance with a query, even in the absence of exact textual matches. For example, a query for ``machine learning'' can retrieve nodes related to ``artificial intelligence'' or ``neural networks'' by calculating the similarity between their vector representations. This method is particularly effective for queries requiring semantic understanding and similarity-based matching.

In addition to query vectors, the system can utilize vectors of hypothetical answers~\cite{gao2022precise} to enhance retrieval capabilities. This approach identifies information closely related to potential answers, significantly expanding the scope of retrieved data. However, managing the resulting large volume of information poses a challenge, which we address in Section~\ref{sec:Post Processing}.

Vector retrieval also supports clustering and classification tasks within the PKG, enabling the grouping of similar nodes and the identification of patterns across the graph. For instance, in social media analysis, vector-based clustering can identify trending topics or clusters of related content, offering insights into user interests and emerging discussions. In scientific research, it can map relationships between research papers, revealing interdisciplinary connections even in the absence of direct citations. This capability accelerates knowledge discovery and fosters collaboration across fields.

In summary, vector retrieval enhances semantic understanding and enables the discovery of intricate connections between entities through high-dimensional vector representations. By incorporating vectors for queries, entities, and hypothetical answers, it provides a powerful mechanism for uncovering patterns and trends. While effective, the method requires careful management of retrieved data to ensure efficiency and relevance. Overall, vector retrieval significantly advances data analysis and insight generation across diverse domains.t is crucial. Ultimately, vector retrieval advances data analysis and insight generation across various fields.

\begin{figure}[ht]
    \centering
    \subfigure[A subgraph of PKG]{
        \includegraphics[width=0.4\textwidth]{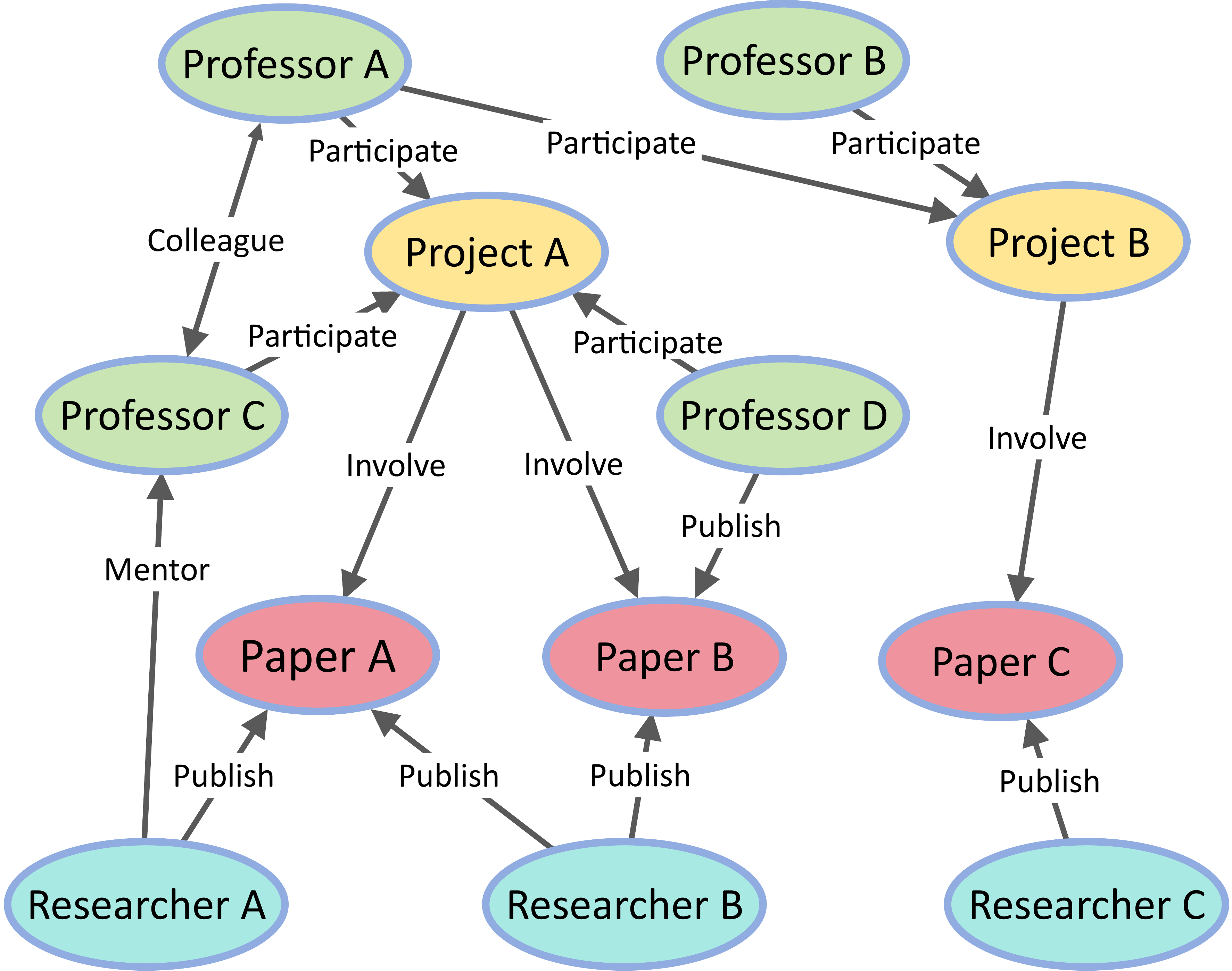}
        \label{fig:metapath sub1}
    }
    \subfigure[``professor-project-professor"]{
        \includegraphics[width=0.2\textwidth]{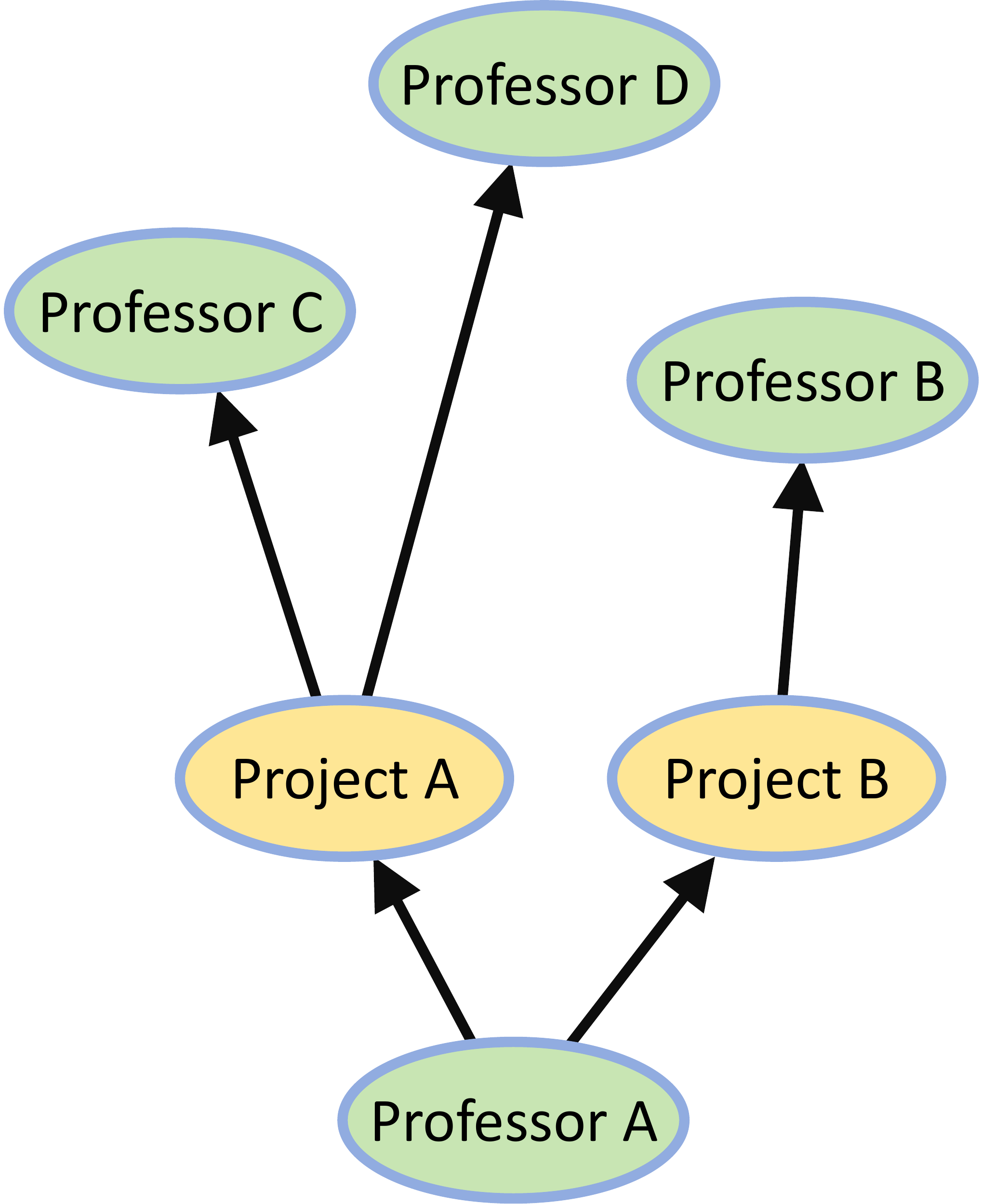}
        \label{fig:metapath sub2}
    }
    \subfigure[``project-paper-researcher'']{
        \includegraphics[width=0.2\textwidth]{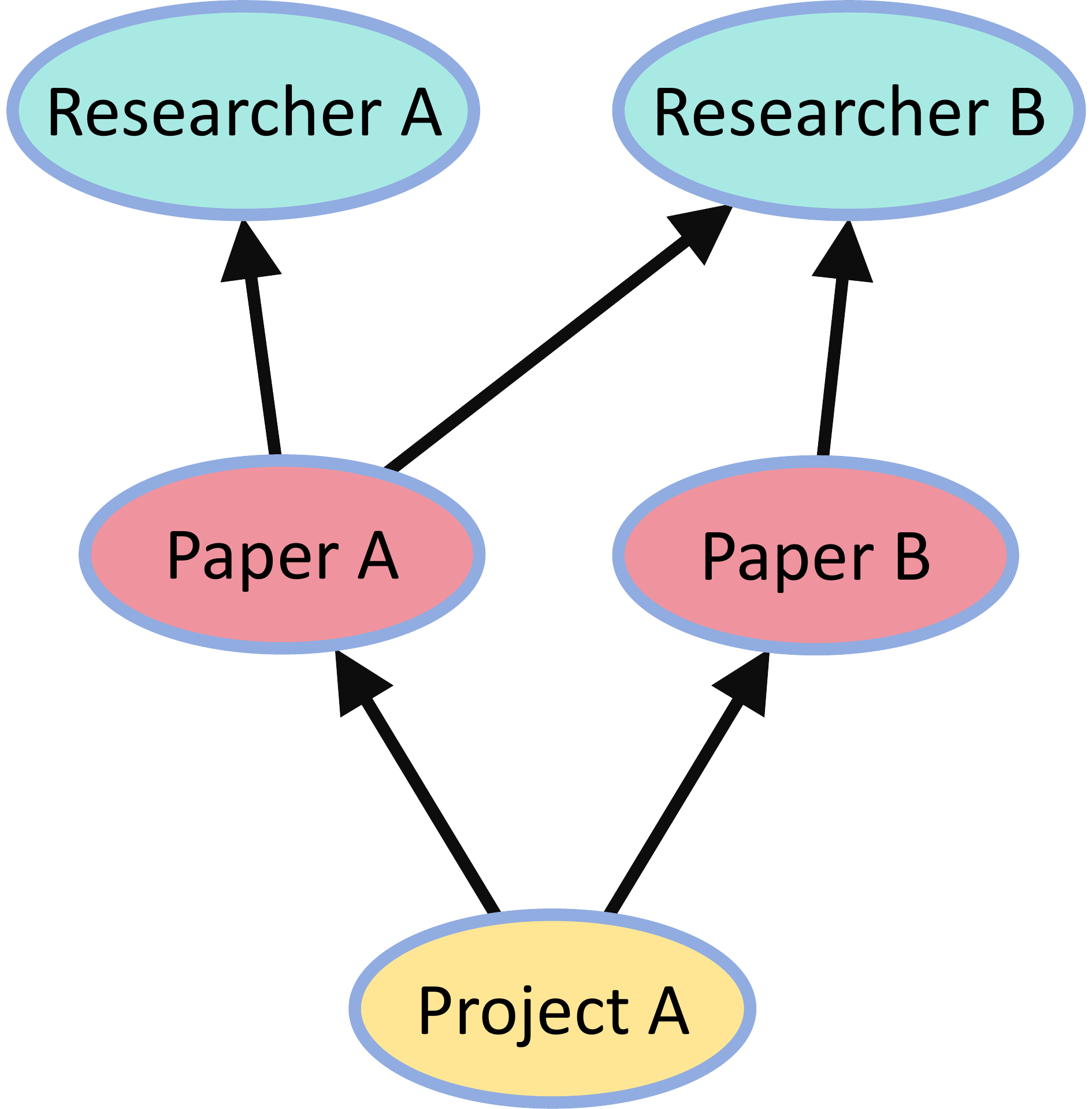}
        \label{fig:metapath sub3}
    }
    \caption{Meta-path Retriever. When using meta-paths, we only care about the node chain but not the relations between them. After we obtain the node sequences, we can analyze the connections among different entities. (b) and (c) illustrate various meta-paths highlighting different relationships and interactions within subgraph (a).}
    \label{fig:metapath}
\end{figure}

\subsubsection{Meta-path Retrieval}\label{sec:metapath}

In traditional KGs, meta-paths are a core feature that enables the exploration of relationships between entities through predefined paths. A meta-path is a sequence of nodes and edges in a knowledge graph that defines a specific relational pathway, allowing for the discovery of complex and multi-hop connections. One can get much more complex relations through meta-paths than normal methods. For example, a meta-path such as ``professor-project-professor'' can identify all professors who have collaborated with a specific professor through shared projects. This capability is unique to meta-path retrieval, as it uncovers deeper and more nuanced relationships than traditional methods like vector-based search, which primarily focus on semantic similarity.

However, integrating meta-paths of knowledge graphs with LLMs presents challenges~\cite{chen2023zero}. On one hand, the computational complexity of dynamically constructing and traversing meta-paths during query processing can lead to high latency, limiting their practical application. On the other hand, although large language models excel in processing natural language, there are still technical barriers to integrating them with knowledge graphs, especially when dealing with meta-paths.

To address these challenges, we propose an innovative retrieval method that reduces the complexity and computational cost of meta-path search. Our approach involves pre-constructing meta-paths of lengths less than a predefined value $n$ and storing them as attributes within the nodes of the PKG. This pre-processing step allows for efficient retrieval of relevant meta-paths during query execution. To further optimize the search process, we integrate a lightweight model that dynamically identifies the most relevant meta-paths for a given query. This model, which is computationally efficient, analyzes the query context and selects the appropriate meta-paths, enabling rapid multi-hop exploration without the need for extensive on-the-fly computation.

For example, in a research collaboration network, meta-paths such as ``project-paper-researcher'' can be pre-stored within node attributes. When a query is issued, the system quickly identifies researchers relevant to specific projects by traversing these pre-constructed paths. This approach not only reduces query latency but also enhances the system's ability to uncover intricate relationships that would otherwise require significant computational resources.

By employing this method, we streamline the meta-path retrieval process, enabling efficient and scalable multi-hop exploration of relationships within the PKG. The pre-storage of meta-paths minimizes computational overhead, while the lightweight model ensures dynamic and context-aware path selection. This innovation significantly enhances the system's ability to handle complex queries, making it particularly valuable for applications requiring deep relational reasoning, such as academic collaboration analysis, social network exploration, and biomedical knowledge discovery.

In summary, our meta-path retrieval method addresses the limitations of traditional approaches by combining pre-constructed meta-paths with a lightweight, context-aware model. This approach not only improves retrieval efficiency but also opens new avenues for exploring complex data relationships, enabling more insightful analyses and applications across diverse domains.

\subsubsection{Post Processing}\label{sec:Post Processing}

After retrieving information using the aforementioned methods, the next step involves integrating and re-ranking the data to identify the most relevant information for the query. This post-processing phase is critical for ensuring the accuracy, coherence, and comprehensiveness of the final output.

The integration process begins by merging results from the three retrieval methods: Regular Expression Retrieval, Vector Retrieval, and Meta-path Retrieval. Each method contributes unique strengths: Regular Expression Retrieval provides precise pattern-based matches, Vector Retrieval enhances semantic similarity by identifying contextually related information, and Meta-path Retrieval uncovers complex relational pathways between entities. For example, in a research database query seeking collaborations between researchers on specific topics, Regular Expression Retrieval can identify exact matches of researcher names or keywords. Vector Retrieval can then expand the scope by finding semantically related topics, even when different terminology is used. Finally, Meta-path Retrieval can trace indirect collaborations through shared projects or co-authorship networks, revealing deeper connections that might otherwise remain hidden.

Once integrated, the re-ranking process~\cite{sachan2022improving,sun2023chatgpt} prioritizes the results based on their relevance to the query. This involves scoring each piece of information using criteria such as frequency of occurrence, semantic relevance, and the strength of relationships identified through meta-paths. Additionally, we leverage the capabilities of large language models (LLMs) to evaluate the relevance and accuracy of the content. By employing LLMs, we can dynamically assess the quality of retrieved information and refine the ranking based on contextual understanding. Furthermore, LLMs can learn from historical query data, identifying patterns and improving the re-ranking process over time. This adaptive approach allows the system to predict which types of information are most likely to be relevant for similar queries in the future, enhancing retrieval performance.

In conclusion, the post-processing phase plays a pivotal role in refining the retrieved data, ensuring that the final output is both comprehensive and highly relevant to the query's requirements. By integrating multiple retrieval methods and leveraging LLMs for re-ranking, the system delivers precise and contextually appropriate results, enabling the generation of accurate and insightful responses to user queries.

\section{Experiment}

In Section~\ref{sec:datasets}, we provide a detailed description of the datasets utilized in our experiments. Section~\ref{sec:models} outlines the models employed and Section~\ref{sec:settings} discusses the experimental settings. Section~\ref{sec:performance} presents the overall performance evaluation, encompassing a variety of evaluation metrics. Section~\ref{sec:ablation} conducts an ablation study to analyze the contribution of individual components to the system's performance. Section~\ref{sec:analysis} offers further analysis, exploring the impact of model size and retrieval methods on the results. Finally, Section~\ref{sec:case study} presents a case study, illustrating the practical application and effectiveness of our approach through specific examples.

\begin{table*}[htp]
    \centering
    \renewcommand{\arraystretch}{1.2}
    \begin{tabular}{c|c|>{\centering\arraybackslash}p{1cm}>{\centering\arraybackslash}p{1cm}>{\centering\arraybackslash}p{1cm}>{\centering\arraybackslash}p{0.75cm}>{\centering\arraybackslash}p{1.5cm}>{\centering\arraybackslash}p{0.75cm}>{\centering\arraybackslash}p{1cm}|cc}
    \hline
    \multirow{2}{*}{Model} & \multirow{2}{*}{Setting} & \multicolumn{7}{c|}{Open Compass } & \multicolumn{2}{c}{MultiHop-RAG } \\
    \cline{3-11} 
    & & MMLU & AGIEval & NQ & CSQA & OpenBookQA & NLI & COPA & Inference & Temporal  \\
    \hline
    \multirow{4}{*}{GPT2} 
    & LLM-Base & 27.3 & 20.5 & 3.6 & 60.2 & 73.0 & 20.3 & 67.0 & 15.3 & 5.6    \\
    & LLM-VDB & 50.3 & \textbf{22.3} & 18.7 & 65.3 & 85.2 & 20.3 & \textbf{68.0} & 63.2 & 21.3  \\
    & LLM-KG & 44.6 & 19.8 & 18.3 & 66.9 & 80.3 & 20.5 & 67.0 & 59.4 & 20.1    \\
    & LLM-PKG (Ours) & \textbf{52.7} & 20.6 & \textbf{19.5} & \textbf{70.1} & \textbf{86.6} & \textbf{20.3} & \textbf{68.0} & \textbf{70.4} & \textbf{22.5}    \\
    \hline
    \multirow{4}{*}{LLaMA-2-7b} 
    & LLM-Base & 45.9 & 40.5 & 19.6 & 66.5 & 58.4 & 32.3 & 67.0 & 22.3 & 9.9    \\
    & LLM-VDB & 53.2 & 45.5 & 22.0 & 70.3 & 79.5 & 33.1 & 67.0 & 72.6 & 26.7  \\
    & LLM-KG & 50.4 & 42.3 & 22.2 & 69.5 & 72.3 & 32.6 & 67.0 & 75.8 & 23.2    \\
    & LLM-PKG (Ours) & \textbf{61.4} & \textbf{48.9} & \textbf{23.1} & \textbf{75.7} & \textbf{85.3} & \textbf{33.5} & \textbf{69.0} & \textbf{82.3} & \textbf{28.9}    \\
    \hline
    \multirow{4}{*}{Phi3-1b} 
    & LLM-Base & 44.3 & 45.1 & 1.9 & 58.3 & 68.4 & 36.5 & \textbf{70.0} & 19.3 & 7.3\\
    & LLM-VDB & \textbf{56.3} & 45.3 & 11.8 & 62.4 & 84.2 & 37.6 & 68.0 & 66.3 & 20.6  \\
    & LLM-KG & 48.6 & 45.5 & 9.7 & 62.1 & 79.3 & 37.3 & \textbf{70.0} & 65.2 & 18.9    \\
    & LLM-PKG (Ours) & \textbf{56.3} & \textbf{45.8} & \textbf{13.6} & \textbf{65.3} & \textbf{86.3} & \textbf{37.8} & \textbf{70.0} & \textbf{88.3} & \textbf{26.8}    \\
    \hline
    \multirow{4}{*}{Qwen2.5-7b} 
    & LLM-Base & 57.8 & 40.5 & 14.2 & 67.5 & 84.4 & 54.9 & 88.0 & 20.5 & 11.8    \\
    & LLM-VDB & 58.2 & 42.6 & 17.8 & 78.3 & 90.3 & 61.3 & 90.0 & 70.1 & 32.3  \\
    & LLM-KG & 58.0 & 40.8 & 17.6 & 77.0 & 86.9 & 63.5 & 89.0 & 65.3 & 28.4    \\
    & LLM-PKG (Ours) & \textbf{65.8} & \textbf{47.3} & \textbf{20.3} & \textbf{78.6} & \textbf{92.2} & \textbf{66.4} & \textbf{91.0} & \textbf{90.0} & \textbf{35.3}    \\
    \hline
    \multirow{4}{*}{ChatGLM3-6B} 
    & LLM-Base & 51.9 & 47.4 & 7.5 & 70.3 & 79.4 & 40.0 & 89.0 & 23.5 & 10.6    \\
    & LLM-VDB & 57.4 & 47.4 & 19.3 & 78.4 & 86.3 & 53.2 & 90.0 & 73.4 & 32.6  \\
    & LLM-KG & 50.8 & 47.0 & 19.8 & 76.3 & 86.0 & 55.6 & 90.0 & 75.6 & 31.1    \\
    & LLM-PKG (Ours) & \textbf{59.7} & \textbf{47.6} & \textbf{22.7} & \textbf{79.0} & \textbf{88.7} & \textbf{63.2} & \textbf{91.0} & \textbf{89.3} & \textbf{33.4}    \\
    \hline
    \end{tabular}%
    \caption{Performance comparison of different settings on various models across two datasets, using seven indicators for Open Compass and two for MultiHop-RAG. The best performances are indicated in bold font.}
    \label{tab:overall}
    \vspace{-20pt}
\end{table*}

\subsection{Datasets}\label{sec:datasets}
We selected Open Compass~\cite{buitrago2019open} and MultiHop-RAG~\cite{tang2024multihop}, two datasets comprising approximately one million tokens—equivalent to the text of about ten novels—to represent the vast and diverse corpora encountered in real-world scenarios. Open Compass emphasizes user-driven interactions, providing a rich foundation for evaluating models' ability to handle conversational and context-aware tasks. In contrast, MultiHop-RAG focuses on structured, multi-hop reasoning, challenging models to synthesize information across multiple documents and perform complex inference. Together, these datasets offer complementary evaluation frameworks, enabling a comprehensive assessment of our proposed method across a wide range of contexts and tasks, from conversational understanding to advanced reasoning and information synthesis.

\subsubsection{Open Compass}

Open Compass is a specialized dataset designed to evaluate the performance of language models across a wide range of natural language processing (NLP) tasks. It includes diverse user-generated content, such as questions and responses, which reflect real-world interactions. This dataset is particularly valuable for assessing models' comprehension and generation capabilities in practical scenarios. Open Compass is structured to test various aspects of language understanding, including:

\begin{itemize}
    \item MMLU (Massive Multitask Language Understanding)~\cite{hendrycks2020measuring}: Measures the model's ability to perform well across a broad spectrum of tasks, including humanities, STEM, and social sciences.
    \item AGIEval (AI General Intelligence Evaluation)~\cite{zhong2023agieval}: Evaluates the model's performance on tasks that require general intelligence, such as logical reasoning and problem-solving.
    \item NQ (Natural Questions)~\cite{hasan2024nativqa}: Tests the model's ability to answer fact-based questions by retrieving relevant information from a large corpus.
    \item CSQA (Commonsense Question Answering)~\cite{talmor2018commonsenseqa}: Assesses the model's ability to leverage commonsense knowledge to answer questions that require reasoning beyond explicit facts.
    \item OpenBookQA~\cite{alkhaldi2023studies}: Evaluates the model's ability to answer questions by combining explicit knowledge with reasoning, simulating open-book exams.
    \item NLI (Natural Language Inference)~\cite{nie2019adversarial}: Tests the model's ability to understand and infer relationships between sentences, such as entailment and contradiction.
    \item COPA (Choice of Plausible Alternatives)~\cite{huang2024copa}: Measures the model's ability to choose the most plausible outcome or cause in a given scenario, requiring causal reasoning.
\end{itemize}

By encompassing these diverse tasks, Open Compass provides a comprehensive evaluation framework for assessing the robustness and versatility of language models in real-world applications.

\subsubsection{MultiHop-RAG}

MultiHop-RAG is a benchmark dataset specifically designed for multi-hop reasoning tasks, requiring models to connect information across multiple documents to answer complex queries. It comprises an extensive collection of news articles published between September 2013 and December 2023, covering categories such as entertainment, business, sports, technology, health, and science. The dataset is structured to evaluate models' ability to synthesize information from disparate sources and generate coherent, contextually appropriate responses. Key features of MultiHop-RAG include:

\begin{itemize}
    \item Inference Query: Requires the model to perform multi-hop reasoning by connecting information from different articles to deduce the correct answer. This tests the model's ability to integrate and reason over multiple pieces of information.
    \item Temporal Query: Evaluates the model's ability to analyze and utilize temporal information within the retrieved data, such as identifying the chronological order of events or understanding time-sensitive contexts.
\end{itemize}

MultiHop-RAG is particularly challenging due to its emphasis on multi-hop reasoning and temporal understanding, which are critical for tasks requiring deep contextual analysis and synthesis of information from multiple sources. The dataset's complexity makes it an ideal benchmark for evaluating advanced retrieval and reasoning capabilities in language models.

\subsection{Models}\label{sec:models}
We adopt the following common Open-source LLMs as base models for comparison different settings in Section~\ref{sec:settings}:

\begin{itemize}
    \item \textbf{GPT}~\cite{radford2018improving} is a groundbreaking language model that uses a transformer architecture to generate coherent and contextually relevant text. We selecte GPT-2 for our work as it is the latest open-source model in the GPT family.
    
    \item \textbf{LLaMA}~\cite{touvron2023llama} is a series of models designed for efficient language processing. Notably, LLaMA-2-7b excels in generating and understanding text, demonstrating high performance across various tasks.
    
    \item \textbf{Phi}~\cite{abdin2024phi} introduces an innovative approach to language modeling by combining transformer architectures with novel neural network designs, enhancing both understanding and generation capabilities.
    
    \item \textbf{ChatGLM}~\cite{glm2024chatglm} is a conversational AI model optimized for interactive dialogue. Its sophisticated architecture improves context understanding and provides informative responses in real-time interactions.
    
    \item \textbf{Qwen}~\cite{bai2023qwen} comprises a range of models with varying parameter sizes. In our experiments, we utilized different models from the Qwen2.5 family, including 0.5B, 1.5B, 3B, and 7B, to explore their performance across tasks.
    
\end{itemize}

\subsection{Settings}\label{sec:settings}

In our analysis, we aim to evaluate the performance and capabilities of various models under different retrieval conditions. Specifically, we investigate four distinct configurations, as follows:

\subsubsection{LLM-Base (Only LLM)}

This approach employs a standard language model to address user queries without any additional context or retrieval mechanisms, which means that all the information in the answers comes entirely from the model itself.

\subsubsection{LLM-VDB (LLM with Vector Database RAG)}

In this setup, we enhance the language model's capabilities by integrating a retrieval-augmented generation (RAG) approach that utilizes a vector database~\cite{jing2024large} to provide relevant context for answering queries. In our setup, we use Elasticsearch~\cite{elasticsearch2018elasticsearch} as the vector database for retrieval-augmented generation. After setting up an Elasticsearch cluster and indexing documents with vector embeddings, queries are processed by searching the vector space to find the most relevant context.

\subsubsection{LLM-KG (LLM with traditional Knowledge Graph)}

A traditional KG is employed as the retriever. KGs represent information in a structured format, using nodes and edges to capture relationships between entities.In our setup, We use LightRAG~\cite{guo2024lightrag} as the KG retriever. We need first construct a KGintegrate LightRAG with KG by setting up the model to query nodes and edges, retrieving relevant information to enhance the language model's responses. This involves configuring LightRAG to interact with your existing KG structure.

\subsubsection{LLM-PKG (LLM with Pseudo-Knowledge Graph)}

Here, we enhance the model by integrating a PKG, as detailed in Section~\ref{sec:methodology}. This addition enables the system to access relevant data dynamically and enriches the response quality.

\subsection{Overall Performance}\label{sec:performance}

We evaluate the proposed Pseudo-Knowledge Graph (PKG) framework against various baseline models using multiple metrics, with the overall results presented in Table~\ref{tab:overall}. The analysis reveals the following key insights:

For the baseline method, the LLM models perform well on Exams (\emph{i.e.}, MMLU and AGIEval), and Reasoning (\emph{i.e.}, NLI and COPA), but not for Knowledge (\emph{i.e.}, NQ and CSQA) and Understanding (\emph{i.e.},  OpenBookQA). These models struggle with knowledge-based tasks, which demand a deeper comprehension of scientific facts and the ability to connect disparate pieces of information. This discrepancy arises because LLMs tend to perform well with materials similar to their training data. When encountering unfamiliar information, their understanding is limited. By incorporating RAG, LLMs can access external knowledge sources, which enhances their performance on knowledge-based tasks. This access allows them to pull in relevant information from vast databases, improving their ability to answer fact-based questions. Similarly, integrating KGs with LLMs provides a significant boost. KGs offer structured information and complex relationships between data points, which not only enrich the knowledge base but also improve reasoning and understanding capabilities. LLMs with KGs outperform those with RAG in these areas because KGs provide a richer context and a more nuanced understanding of how different pieces of information interrelate, thereby enhancing the model's ability to interpret and reason through complex scenarios. However, due to the LLMs' limited ability to understand structured data, KGs do not perform as well as vector database-based RAG for tasks that do not require strong logical reasoning.

Our proposed PKG maintains the best performance on most dataset metrics and shows significant improvements compared to the baseline methods. This superior performance can be attributed to three factors: i) We provide LLMs with a rich information through PKG, leveraging diverse retrieval methods. This diversity results in a wider variety of information types and higher quality data, enhancing the model's ability to understand and generate accurate responses. ii) By retaining original text chunks within the PKG, LLMs can bypass the complexities of processing structured data. This enables the models to better comprehend and interpret the knowledge, as they can work with familiar unstructured text formats. iii) We utilize meta-paths to perform more complex relationship analyses, which significantly enhances our method's performance in understanding and reasoning tasks. This capability allows the model to discern intricate patterns and connections within the data, leading to superior results in these challenging areas.

\subsection{Ablation Study}\label{sec:ablation}

\begin{table}[htp]
    \centering
    \renewcommand{\arraystretch}{1.2}
    \begin{tabular}{c|l|>{\centering\arraybackslash}p{1.5cm}|>{\centering\arraybackslash}p{1.5cm}}
    \hline
    \multirow{2}{*}{Model} & \multicolumn{1}{c|}{\multirow{2}{*}{Setting}} & \multicolumn{2}{c}{Open Compass }  \\
    \cline{3-4} 
    & &  CSQA & OpenBookQA   \\
    \hline

    \multirow{4}{*}{Qwen2.5-7b} 
    & \ LLM & 20.5 & 11.8     \\
    & \ \ + NLP EX & 75.2 & 83.4   \\
    & \ \ + LLM EX & 77.5 & 86.7    \\
    & \ \ + ING TEXT & \textbf{78.6} & \textbf{92.2}     \\
    \hline
    \end{tabular}%
    \caption{Ablation study of various Building and Storage methods in PKG Builder. We show the results on CSQA and OpenbookQA in Open Compass dataset.}
    \label{tab:Ablation1}
    \vspace{-20pt}
\end{table}

\subsubsection{Building and Storage}
Our proposed PKG Builder consists of various components, including: i) traditional NLP-based Extraction (NLP EX): this method utilizes established natural language processing techniques such as tokenization and rule-based named entity recognition; ii) LLM-based Extraction (LLM EX): this approach leverages LLMs like GPT to interpret and extract information, allowing for more nuanced and flexible extraction of information from complex and unstructured text; and iii) In-graph text chunks (ING TEXT): by embedding text segments directly within the PKG, we preserve the complete information from the original text and this helps LLMs better understanding the knowledge. To assess the effectiveness of each component, we perform an ablation study using the CSQA and OpenBookQA datasets on Qwen2.5-7b. These datasets are chosen for its complexity and rich knowledge content, allowing us to analyze the contribution of each part thoroughly.

The results, as shown in Table~\ref{tab:Ablation1}, demonstrate that both traditional NLP and LLM methods significantly enhance the performance of the PKG. Particularly noteworthy is the impact of embedding in-graph text chunks. This approach preserves the full context of the original information, enabling language models to understand knowledge across various scenarios more effectively. By maintaining the integrity of the source material, these in-graph text chunks enable deeper insights and more accurate interpretations. All components of the PKG Builder contribute to constructing a robust PKG, offering significant potential for future retrieval tasks.

\begin{table}[htp]
    \centering
    \renewcommand{\arraystretch}{1.2}
    \begin{tabular}{c|l|>{\centering\arraybackslash}p{1.2cm}|>{\centering\arraybackslash}p{1.2cm}}
    \hline
    \multirow{2}{*}{Model} & \multicolumn{1}{c|}{\multirow{2}{*}{Setting}} & \multicolumn{2}{c}{MultiHop-RAG}  \\
    \cline{3-4} 
    & &  Inference & Temporal   \\
    \hline

    \multirow{4}{*}{Qwen2.5-7b} 
    & \ LLM & 20.5 & 10.6     \\
    & \ \ + REG RE & 60.4 & 25.3   \\
    & \ \ + VEC RE & 75.1 & 31.7    \\
    & \ \ + META-PATH RE & \textbf{90.0} & \textbf{35.3}     \\
    \hline
    \end{tabular}%
    \caption{Ablation study of various Retrieval methods in PKG Retriever. We show the results on Inference and Temporal in MultiHopRAG dataset.}
    \label{tab:Ablation2}
    \vspace{-20pt}
\end{table}

\subsubsection{Retrieval Methods}\label{sec:Ablation Retrieval}
In addition to the building of PKG, we also examine the proposed retrieval methods, including: i) Regular Expression Retrieval (REG RE): this method uses patterns matching to search and retrieve specific information from PKG; ii) Vector Retrieval (VEC RE): by converting text into high-dimensional vectors using techniques like embeddings, this approach allows for semantic search, enabling retrieval based on the meaning rather than exact match; iii) Meta-path Retrieval (META-PATH RE): this technique involves navigating through the meta-paths in PKG to retrieve information, leveraging the relationships between entities. To validate the effectiveness of each retrieval method, we conduct an ablation study on the MultiHop-RAG dataset with Qwen2.5-7b to analyze the contribution of each part. The MultiHop-RAG dataset is chosen because it presents a significant challenge for retrieval technology. It requires not only finding relevant texts but also understanding and reflecting the relations between them. This complexity makes it an ideal test for advanced retrieval systems.

The results, as shown in Table~\ref{tab:Ablation2}, demonstrate that the basic retrieval method, regular matching, provides external world knowledge to the LLMs in addition to their inherent weights. This expanded knowledge significantly enhances the LLMs' performance as tasks require extensive information. However, for inference, regular expression retrieval does not perform effectively. In contrast, vector retrieval and meta-path retrieval show superior performance in most situations. Vector retrieval excels because it offers semantic information by capturing the meaning and context of words and phrases, allowing LLMs to understand and process nuanced language patterns. This method leverages embeddings to match queries with relevant data points, enhancing the model's ability to draw connections and infer meanings based on the semantic similarities of the data. On the other hand, meta-path retrieval provides a structured way to represent relations between entities, which is crucial for reasoning tasks. By outlining relational paths, this method helps LLMs understand complex interactions and dependencies, enabling more accurate inference and deduction. This approach is particularly beneficial for tasks that require understanding the underlying structure of information and drawing logical conclusions from interconnected data points.

Overall, while regular matching serves as a foundational method for expanding knowledge, the combination of vector retrieval and meta-path retrieval offers a more sophisticated and effective approach for enhancing LLMs' reasoning capabilities. Together, these three retrieval methods form a comprehensive PKG retrieval system.

\subsection{Further Analysis}\label{sec:analysis}

\begin{figure}[htp]
    \centering
    \includegraphics[width=\columnwidth]{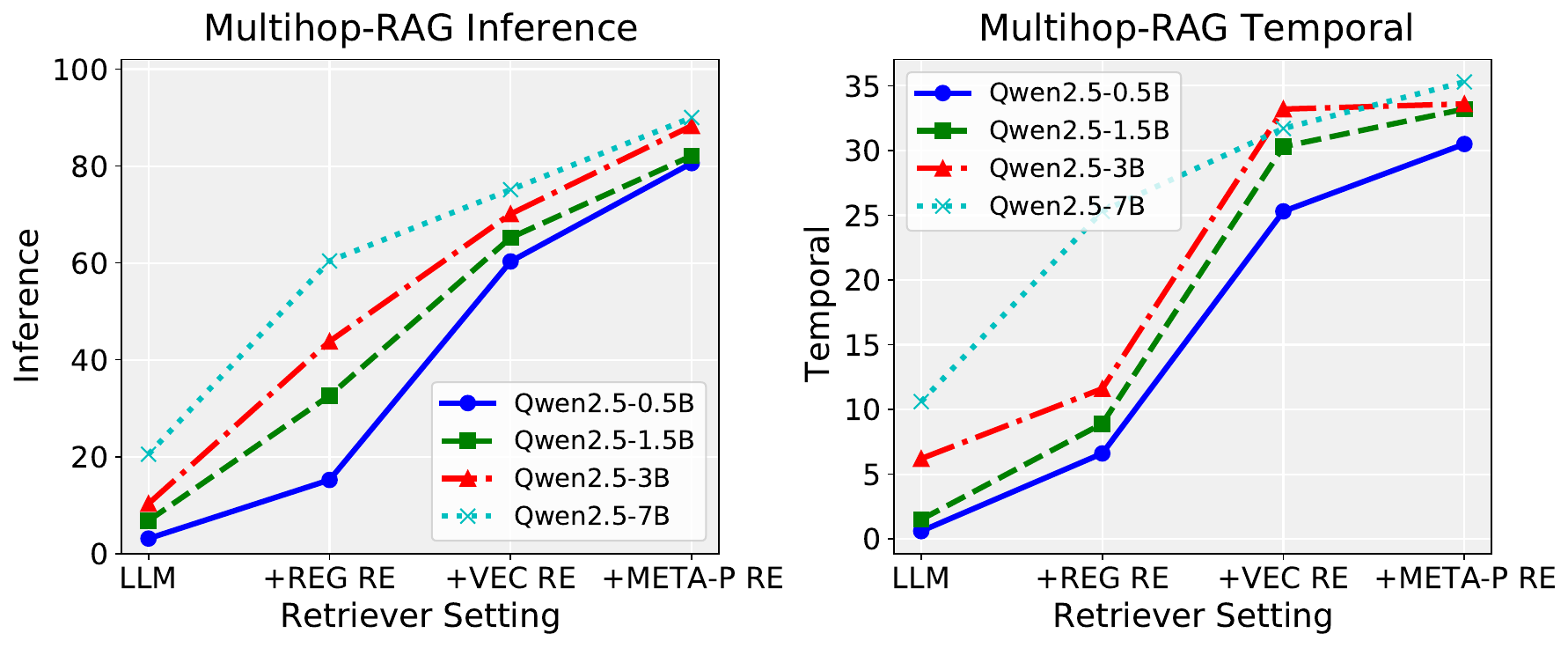}
    \caption{Performance comparison of different retriever settings on Multihop-RAG dataset. The left panel shows inference performance, while the right panel illustrates temporal performance across various configurations (Qwen2.5-0.5B, Qwen2.5-1.5B, Qwen2.5-3B, and Qwen2.5-7B).}
    \label{fig:further}
\end{figure}

After conducting ablation experiments, we aimed to explore the impact of PKG on models of varying sizes. Building on the ablation tests in Section~\ref{sec:Ablation Retrieval} we evaluated the PKG system on different sizes of the Qwen model, specifically 0.5B, 1.5B, 3B, and 7B parameters. We assessed these models' performance on the expanded MultiHop-RAG dataset, with the results illustrated in Figure~\ref{fig:further}. 

The results indicate that when relying solely on the LLM itself, larger models outperform smaller ones due to their greater number of parameters, which encapsulate more information and reasoning capabilities. However, when retrieval results are introduced, the performance of all models improves. Notably, smaller models experience significant enhancement, allowing them to narrow the gap with larger models. This demonstrates the effectiveness of retrieval in augmenting model capabilities, particularly for those with fewer parameters. Among the different retrieval systems, PKG performs the best, surpassing both vector-based knowledge bases and traditional KGs.

\begin{table*}[htp]
    \centering
    \renewcommand{\arraystretch}{1.5}
    \begin{tabularx}{\textwidth}{|l|X|}
    \hline
    \textbf{Question} & \textbf{In the scientific journals published in 2024, what are some emerging biotechnologies that are considered likely to significantly change agricultural practices?} \\
    \hline
    \textbf{Answer 1 (LLM-Base)}  & It's difficult to provide information about specific journal articles or publications from 2024. However, I can give you an overview of emerging biotechnologies that could potentially impact agricultural practices. Some potential biotechnologies that may influence agricultural practices include:
    \textbf{Gene editing technologies such as CRISPR/Cas9[...]Synthetic biology[...]Microbiome engineering[...]Precision agriculture technologies[...]}\\

    \hline
    \textbf{Answer 2 (LLM-VDB)} & Based on the article from the scientific journals published in 2024, there are several emerging biotechnologies that are likely to significantly change agricultural practices:
    \textbf{PrimeRoot System for Precise DNA Insertion[...]PASTE Technology for Genome Insertion[...]dCas9-SSAP for Genomic Knock-in[...]}
    These technologies have the potential to greatly impact agricultural practices by enabling precise genetic modifications in plants, which can lead to crops with improved resistance to diseases, enhanced nutritional profiles, and better adaptability to environmental conditions. The ability to insert large DNA sequences accurately and efficiently is a significant advancement in the field of plant biotechnology and could lead to the development of new crop varieties that are more productive and resilient.\\
    \hline
    \textbf{Answer 3 (LLM-KG)} & In the scientific journals published in 2024, some emerging biotechnologies that are considered likely to significantly change agricultural practices include:
    \textbf{Large DNA Insertion Technologies[...]Drag-and-Drop Genome Insertion[...]dCas9-based Gene Editing for Cleavage-Free Genomic Knock-In[...]Cell Atlas Mapping[...]Super-Resolution Microscopy[...]3D Printing of Nanomaterials[...]}
    These biotechnologies hold the potential to revolutionize agricultural practices by enabling precise genetic modifications in crops, enhancing our understanding of cellular processes, and providing new tools for materials design and imaging techniques.\\
    \hline
    \textbf{Answer 4 (LLM-PKG)} & In 2024, several emerging biotechnologies have been highlighted in scientific journals for their potential to significantly transform agricultural practices. These technologies focus on precise genetic modifications and advanced imaging techniques, which can enhance crop traits and improve our understanding of cellular processes. Here’s an integrated overview:
    \textbf{Large DNA Insertion Technologies[...]CRISPR-Directed Integrases (PASTE)[...]dCas9-based Gene Editing (dCas9-SSAP)[...]Cell Atlas Mapping[...]Super-Resolution Microscopy[...]}
    These biotechnologies offer groundbreaking methods for modifying plant genomes with precision and efficiency. They hold the potential to develop crops with enhanced disease resistance, improved nutritional content, and better environmental adaptability, ultimately leading to significant advancements in agricultural productivity and food security. \\
    \hline
    \end{tabularx}%
    \caption{Example question for the biotechnologies article dataset, with generated answers from LLM-Base, LLM-VDB, LLM-KG and LLM-KG as shown in Section~\ref{sec:settings}.}
    \label{tab:case study}
    \vspace{-20pt}
\end{table*}

\subsection{Case Study}\label{sec:case study}

LLMs have been shown to be good evaluators of natural language generation, achieving state-of-the-art or competitive results compared against human judgements~\cite{wang2024feb4rag}. In this section, we will explore different retrieval systems based on a given question, as shown in Table~\ref{tab:case study}. We provide different kinds of retrievers with a large knowledge base, containing information needed to answer the question and others irrelevant to the question. We want to see if the retrieval system can get the information accurately and precisely from the knowledge base and process them for LLMs to use. Table~\ref{tab:case study} shows an example of LLM-generated assessment.

For LLM itself, it can provide general insights based on its training data. It identifies broad categories of emerging biotechnologies, such as CRISPR and synthetic biology, which have been relevant for years. However, the response may include information that is outdated or speculative, leading to hallucinations. It lacks specificity and does not reflect the most current developments from 2024. LLM with Vector Database RAG uses retrieval to supplement the LLM's responses with more recent and specific data. It identifies precise technologies like the PrimeRoot System and PASTE Technology. While it provides accurate information, it may still miss broader context or additional relevant advancements, focusing narrowly on certain technologies. LLM with KG Integrates a KG allows for a structured and interconnected understanding of the topic. This approach can highlight relationships between technologies and their potential impacts but may still lack depth in explaining each technology's specific mechanisms and applications, potentially leading to less comprehensive answers. The LLM with PKG approach combines the strengths of retrieval methods and knowledge graphs to offer detailed and structured insights. It covers a wide range of technologies and their applications, providing a comprehensive overview. This method excels because it integrates recent, specific data with structured knowledge, using natural language and node-relation chains through meta-paths. As a result, it delivers well-rounded and accurate answers.

After obtaining the retrieval results and generating the answers, we will use GPT-4o~\cite{achiam2023gpt} to assess the quality of the answers generated by language models using these results. The evaluation criteria will include:
\begin{itemize}
    \item \textbf{Accuracy}: The correctness of the information provided in the answer.
    \item \textbf{Coherence}: The logical flow and clarity of the answer.
    \item \textbf{Comprehensiveness}: Whether the answer covers as many aspects of the question as necessary.
\end{itemize}

By applying these evaluation criteria, we demonstrate the effectiveness of different retrieval systems in supporting high-quality answer generation. The LLM-PKG approach (Answer 4) outperforms other methods across all metrics. In terms of accuracy, it clearly identifies specific biotechnologies, such as the PrimeRoot system, PASTE, and dCas9-SSAP, and provides detailed explanations of their mechanisms and potential impacts on agriculture. This contrasts with LLM-Base (Answer 1), which discusses unrelated biotechnologies, and LLM-VDB (Answer 2) and LLM-KG (Answer 3), which focus on general scientific advancements rather than specifically addressing emerging agricultural technologies.

In terms of coherence, LLM-PKG delivers a logically structured response with clear headings and concise explanations, making it easy to follow. In contrast, LLM-Base lacks coherence due to its disjointed discussion of unrelated topics, while LLM-VDB and LLM-KG fail to directly address the original question, resulting in a fragmented flow.

Finally, in terms of comprehensiveness, LLM-PKG stands out by outlining multiple emerging biotechnologies, explaining their applications in agriculture, and including details on both genetic modification and advanced imaging techniques. While LLM-VDB and LLM-KG provide more detailed information than LLM-Base, they still lack the broader context and organizational clarity of LLM-PKG.

\section{Conclusion And Future Work}

In this paper, we introduce the Pseudo-Knowledge Graph (PKG), a Retrieval-Augmented Generation (RAG) framework designed to address the limitations of traditional RAG systems, particularly in managing complex relationships within large-scale knowledge bases. PKG integrates both structured data (knowledge graphs) and unstructured data (in-graph text chunks) to enhance the retrieval capabilities of large language models (LLMs). By preserving natural language text within the graph structure, PKG enables LLMs to process and interpret retrieved information more effectively, overcoming their inherent limitations in handling purely structured data. To seamlessly integrate PKG with LLMs, we develop a suite of advanced retrieval methods, including regular expression retrieval, graph-based vector retrieval, and meta-path retrieval. These methods collectively improve both the semantic understanding and efficiency of information retrieval, ensuring that the retrieved results align closely with the LLM's comprehension and contextual awareness. Extensive experiments across multiple datasets and frameworks demonstrate that PKG outperforms several competitive baseline models and mainstream RAG approaches, particularly in tasks requiring complex reasoning and multi-hop retrieval.

Looking ahead, we plan to extend PKG in several directions to further enhance its capabilities and applicability:
\begin{itemize}
    \item Multi-Turn Conversations: We aim to adapt PKG to support multi-turn conversational interactions, enabling more dynamic and context-aware dialogues with users. This will involve developing mechanisms to maintain context across multiple queries and responses.
    
    \item Scalability and Efficiency: As knowledge bases continue to grow, we will focus on optimizing PKG's scalability and computational efficiency, particularly for real-time applications and large-scale deployments.
    
    \item Interactive Knowledge Exploration: We envision extending PKG to support interactive knowledge exploration, allowing users to navigate complex knowledge graphs intuitively and extract insights through natural language queries.
    
\end{itemize}

By pursuing these directions, we aim to further advance the capabilities of PKG, making it a versatile and powerful tool for enhancing LLMs in both general and domain-specific applications.




\clearpage

\bibliographystyle{ACM-Reference-Format}
\bibliography{long.bib}

\end{CJK*}
\end{document}